\documentstyle[aps,preprint]{revtex}

\def\where{\mathop{\rm where}}
\def\and{\mathop{\rm and}}

\begin{document}

\title{ The Fractional Quantum Hall Effect on a Lattice.}
\author{ F.F. Assaad$^{(1)}$ and  S. Biskamp$^{(2)}$ \\
	 $^{(1)}$ ISSP, The University of Tokyo, Roppongi, 
         Minato-ku, Tokyo 106, Japan.\\
         $^{(2)}$ Physikalisches Institut, Universit\"at W\"urzburg 
         97074 W\"urzburg, FRG.}

\maketitle

\begin{abstract}
Starting from the Hofstadter butterfly, we define lattice versions of Landau 
levels as well as a continuum limit which ensures that they
scale to continuum Landau levels.
By including a next-neighbor repulsive interaction and projecting 
onto the lowest lattice Landau level, we show that 
incompressible ground states exist at filling fractions, 
$\nu = 1/3, 2/5 $ and $3/7$.
Already for values of $l_0/a \sim 2$ where $l_0$ ($a$) is the magnetic length
(lattice constant), the lattice version of the $\nu = 1/3$ state reproduces
with nearly perfect accuracy the the continuum Laughlin state. 
The numerical data strongly suggests that at odd filling fractions of the
lowest lattice Landau level, the lattice constant is an irrelevant
length scale. 
We find a new
relation between the hierarchy of incompressible states and the
self-similar structure of the Hofstadter butterfly.
\end{abstract}

\newpage
There has been a considerable amount of numerical work on the 
fractional quantum Hall effect \cite{Prange} (FQHE) for continuum models
on planar or spherical geometries. 
The question we wish to address in this paper is: can one observe
the FQHE on a lattice and if so,
how does it compare to continuum results? 
Since on the lattice rotational invariance is broken, the 
pseudopotential \cite{Haldane1}  approach to the FQHE is not applicable 
and there is a priori no reason to  expect that
ground states at certain filling fractions of Hofstadter \cite{Hofst} bands are
characterized by a specific power law decay of the density-density
correlation $g(|\vec{r}| )$ for small values of $|\vec{r}|$.
On the other hand there is an efficient way to
construct incompressible quantum liquids (IQL) states
which implicitly assumes the existence of Laughlin states \cite{Laughlin} 
on a lattice:
the parton approach \cite{Wen,Jain}.
The Hamiltonian which has parton wave functions as mean-field solutions
can be derived from first principles only on a lattice \cite{Wen}.
Hence, this approach requires that
properties of the Laughlin states survive discretization of the
plane.
We show that even for small lattice sizes there exist
incompressible states at odd denominator filling fractions of the lowest
Hofstadter level if it has the Landau level degeneracy. These states
reproduce with remarkable accuracy the relevant properties of the
Laughlin states.

Our starting point are fully spin-polarized electrons on a
square lattice in a magnetic field:
\begin{equation}
\label{Ham1}
     H_0  =  -t \sum_{<\vec{i},\vec{j}>}
       c_{\vec{i}}^{\dagger} c_{\vec{j}}
       \exp \left(  \frac{2\pi i}{\Phi_0} \int_{\vec{i}}^{\vec{j}}
       \vec{A} d\vec{l} \right).
\end{equation}
Here, the sum runs over next-neighbors, $c_{\vec{i}}^{\dagger}$ 
creates a spinless electron on site
$\vec{i}$,  $n_{\vec{i} } =  c_{\vec{i}}^{\dagger}  c_{\vec{i}}$ and in the
Landau gauge $ \vec{A}(\vec{x}) = -B(x_2,0,0)$. 
The square lattice lies in the $\vec{e}_1$, $ \vec{e}_2$
plane and $\Phi_0$ is the flux quantum.
In order to put the system on a torus,
we impose the boundary conditions:   $c_{\vec{i} + \vec{L} }^{\dagger} =
     \exp \left( \frac{2\pi i}{\Phi_0} \chi_{\vec{L}} (\vec{i})  \right)
     c_{\vec{i}}^{\dagger}$.  
Here, $\vec{L}$ is a vector with norm
equal to the linear size of
the square lattice oriented along the $\vec{e}_1$ or $\vec{e}_2$
direction, and  $ \vec{A}(\vec{x} + \vec{L} ) $ $=\vec{A}(\vec{x}) +
\vec{\bigtriangledown} \chi_{\vec{L}} (\vec{x})$. 
To obtain a single-valued wave function the total flux traversing
the lattice has to be an integer multiple of the flux quantum
\cite{Fradkin}. The spectrum of the above Hamiltonian yields
the Hofstadter butterfly \cite{Hofst}.
To make contact with the corresponding  continuum Hamiltonian, 
we define the field operators 
$  c_{\vec{i}}^{\dagger}  =  a \int d\vec{r}  \delta( \vec{r} - \vec{R_i})
\Psi^{\dagger} (\vec{r} ) $.
Expanding the phase factors in equation (\ref{Ham1}) and the
delta-function up to second order in the
lattice constant, one obtains up to a constant term:
$ H_0 \rightarrow
    \frac{t a^2}{\hbar^2} \int d\vec{r} \Psi^{\dagger} (\vec{r} ) 
           \left( \vec{P} + \frac{e}{c}  \vec{A} (\vec{r}) \right)^2
\Psi (\vec{r})$.
Setting $ t a^2/\hbar^2 \equiv 1/2M $ where
$M$ denotes a constant mass,  yields the desired continuum limit. 
The continuum Hamiltonian has an energy spectrum described by Landau
levels:
$ E_n =  \hbar \omega_c ( n + 1/2) $ where $ \omega_c = e B/Mc$.  Each
Landau level has a macroscopic degeneracy given by $ N_\Phi = L^2 B/
\Phi_0$ and the magnetic length is given by $l_0 = \left( \frac {2 \pi
B}{\Phi_0} \right)^{-1/2}$.

To achieve such a situation on the lattice,  we set 
\begin{equation}
\label{flux}
	\frac{p}{q}  = \frac { B a^2}{\Phi_0} = \frac{1}{mN} \; \; {\rm for \;  \; a } \;\; N \times N 
\; \; {\rm lattice}.
\end{equation}
Since the number of flux quanta traversing the lattice, $N_\Phi =
N/m$, has to be an
integer, $N$ has to be divisible by $m$.
Before arguing why this choice of $p/q$ is suitable, let us note
that in order to obtain a well defined continuum limit for a fixed value
of $m$, we have
to set $B$ or equivalently $l_0$ to a constant. This 
yields: $a \sim 1/\sqrt{N}$ as well
as $L \sim \sqrt{N}$. Hence, our continuum limit corresponds to a
combined continuum ($a \rightarrow 0$) and thermodynamic ($L
\rightarrow \infty$) limit.  Since we want to keep the
filling fraction $\nu =  N_p/N_\Phi $ constant during the scaling to the
combined continuum and thermodynamic limit, $N_p/N^2   \sim 1/N$. Here,
$N_p$ denotes the number of particles on the lattice. 

For the above choice of $p/q$ and in the Landau gauge, the boundary
conditions are: 
    $ c_{\vec{i} + N\vec{a}_1} = c_{\vec{i}}$  and
    $ c_{\vec{i} + N\vec{a}_2} = \exp\left( -2 \pi i x /m
\right) c_{\vec{i}}$.
Here, $\vec{i} = a\left(x,n \right)$.
We define the partial fourier transform
$\gamma_k^{\dagger, (n)} = \frac{1}{\sqrt{N}}
\sum_{x} e^{-i k x} c_{a(x,n)}^{\dagger}$, 
$ k = 2 \pi n_k/N $ where 
$n_k = 1 \dots N$ to write the Hamiltonian (\ref{Ham1}) as:
\begin{eqnarray}
\label{Ham2}
 H_0 & = &  \sum_k  \sum_n \left( 
	 -2t \cos \left( \frac{2 \pi n }{mN} + k \right)
                         \gamma_k^{\dagger, (n)} \gamma_k^{(n)}
          -t \left( \gamma_k^{\dagger, (n)} \gamma_k^{(n+1)}   + h.c. \right)
                      \right)  \nonumber \\ 
    \gamma_k^{\dagger, (n + N)} &  =  &
   \gamma_{k + k_{\Phi}}^{\dagger, (n)}\; \; \;  \where \; \; \;  k_{\Phi} = 2\pi/m.
\end{eqnarray}
The boundary condition in  the $n$ index of the
$\gamma$ operators  mixes $m$ $k$-sectors. Hence,
the resulting Hamiltonian is block diagonal and contains $N_\Phi =
N/m$ blocks, each of of dimension $mN \times mN$.
We label the blocks with the quantum number $k_0$, and denote 
the $mN\times mN$ matrices by $A(k_0)$.
For this special choice of $p/q$ (\ref{flux}), 
one may show that the matrices $A(k_0)$ are equivalent so that the spectrum
of $A(k_0)$ is independent of $k_0$. Therefore, each energy level of 
the Hamiltonian (\ref{Ham2}) is at least $N_\Phi$-fold degenerate. 
Numerically, one may see that the eigenvalues of one  $A(p)$ matrix
are up to accidental degeneracies at zero energy,  non-degenerate. 
The spectrum of the Hamiltonian 
(\ref{Ham2}) is thus described by $Nm$, $N_\Phi$-fold degenerate energy levels.
We identify those energy levels to lattice Landau levels (lLL).  
To proceed further, we have to diagonalize numerically the  matrices 
$A(k_0)$ so as to obtain the single particle wave functions 
spanning a given lLL.
For more general values of $p/q = \frac{n}{m} N$, the Hamiltonian again 
splits into $N/m$ block diagonal matrices of dimension $mN \times mN$.
For $n>1$ these matrices are not equivalent, and the spectrum
is better described as $mN$, $N_\Phi/n$-fold degenerate bands.  One may
see numerically that as the lattice size is increased, 
$n$ low-lying  adjacent bands merge very rapidly together to form 
a Landau level \cite{Note3}.

Having defined lLL, one may now compare their 
properties with those of the continuum Landau levels.  Consider the
magnetic length which is given in the continuum by:  $l_0 = \left( \frac
{2 \pi B}{\Phi_0} \right)^{-1/2}$. For our choice of  $p/q$
the former equation is equivalent to $ a/l_0 = \sqrt{2\pi/(mN)} $.
In order to obtain an estimate of the magnetic length on the lattice, 
one  notes that for a filled first Landau level (i.e. $\nu = 1$)
\cite{Jancovici}:
\begin{equation}
\label{den-den}
   \left(\frac{N_p}{N}\right)^{-2}
      \sum_{\vec{i}}  < n(\vec{i}) n(\vec{i} + \vec{r}) > 
                   \rightarrow
1 - \exp \left( - \frac{r^2}{2 l_0^2} \right).
\end{equation}
In order to obtain the magnetic length on a finite lattice,  we
have fitted our lattice-data for the density-density correlations 
(lhs of the above equation) to the continuum form
(rhs of the above equation).  
We  consider two  directions: $\vec{r} = r (1,1)$ and 
$\vec{r}   = r (1,0)$. For both directions, and already for small
lattice sizes (i.e. $N = 12$) the data fits the above form extremely well
for all considered values of $m$ \cite{Note2}.
Figure 1 plots the lattice magnetic-length for both considered 
directions  (symbols) as well as its value predicted by the
continuum equations (solid line).  As may be seen, the agreement, 
already for small lattice sizes is very good.  It is to be noted that
the rotational invariance present in the continuum is violated by the presence
of the square lattice.  However, this symmetry is restored 
since as may be seen the discrepancy between the two lattice values of
$l_0$ vanishes rapidly. 
To understand this striking result we have considered the continuum
eigenfunctions of $H_0$ and the corrections due to finite lattice
constant and volume. We find that 
the decay of the eigenfunctions
remains gaussian in any order of perturbation theory.
Similarly, one may check that the cyclotron 
frequency scales rapidly to its continuum value.
Hence, the above defined  lLL, reproduce very accurately
the physics of the continuum Landau levels already on small
lattice sizes. 

We show that IQL states exist on the lattice  by considering a 
next-neighbor density-density
interaction:
\begin{equation}
\label{H_I}
    H_{I}  =   \frac{V}{2} \sum_{<\vec{i},\vec{j}>}
                             :n_{\vec{i} } n_{\vec{j}}: \rightarrow
 \frac{V a^4}{2}
\int d\vec{r} : \rho(\vec{r}) \Delta_{\vec{r}} \rho(\vec{r}):
\end{equation}
Here, $:O:$ denotes the normal ordering of the operator $O$.
A similar Hamiltonian was considered in reference \cite{Canright} in
the context of anyon superconductivity. 
For this continuum interaction (rhs of the above equation), it has been
shown that Laughlin wave function is exact \cite{Trugman}.
Therefore, we should ultimately reproduce the physics 
of the Lauglin wave function. 

In order to carry out the calculations, we project the Hamiltonian
onto the lowest lLL. This projection is justified in the limit 
$B \rightarrow \infty$.
The resulting Hamiltonian  
is an effective one dimensional Hamiltonian
which one may solve numerically by using a standard implementation 
of the  Lanczos method.  The projected Hamiltonian has
the same symmetries (conservation of the one-dimensional total momentum)
as the Hamiltonian considered by  Yoshioka et al. \cite{Yoshioka}.
The scaling of the hopping matrix element, $t$, and interaction
strength, $V$, follow from the continuum limit:  
$V = V_0N^2 $ and $t = t_0N$ where $V_0$ and $t_0$ are constants 
independent of the lattice size. The former follows from the fact
that $a \sim 1/\sqrt{N}$.

For the numerical calculations we chose $m=1$ so that $N=N_\Phi$.
This choice gives the smallest magnetic length, and hence 
lattice effects should be maximized.
The following refers to this special case.
Our numerical analysis is based on the calculation of the
quantity $\Delta E_0(N_p/N) \equiv E_0(N_p/N)  -  N_p \epsilon_1^{(N)}$
where $E_0(N_p/N)$ is the ground state energy of
the projected Hamiltonian for an $N$-site chain with
$N_p$ particles and $\epsilon_1^{(N)}$ is the single particle
energy of the first lLL.
The filling fraction is given by
$\nu = N_p/N$. Since the first lLL
is dispersionless $\Delta
E_0(N_p/N)$ is independent of $t$. Hence only interaction effects
are taken into account. We search for incompressible states by
computing the energy gap to adding a full electron,
\begin{equation}
\label{Gapp}
      \Delta_p(\nu) = \Delta E_0\left(\frac{N_p+1}{N} \right) - 
                      2 \Delta E_0\left(\frac{N_p}{N} \right) +
                      \Delta E_0\left(\frac{N_p-1}{N} \right)
\end{equation}
as well as the energy gap to adding one flux quantum,
\begin{equation}
\label{Gapph}
      \Delta_{\Phi_0}(\nu) = \Delta E_0\left(\frac{N_p}{N-1} \right) - 
                      2 \Delta E_0\left(\frac{N_p}{N} \right) +
                      \Delta E_0\left(\frac{N_p}{N+1} \right).
\end{equation}
Figure 2, plots a summary of our results for various filling fractions
and as a function of the lattice size.  Several points may be made:
i) The gap to adding a full electron 
shows more fluctuations  as a function of lattice size than 
the gap to adding a flux quantum.
$\Delta_{\Phi_0}(\nu =2/5)$ and $\Delta_{\Phi_0}(\nu =1/3)$ show
a very smooth behavior as a function of lattice size. This provides
convincing data for the occurrence of incompressible states at $\nu
=1/3$ and $\nu=2/5$ on the lattice. 
The order of the gaps to adding a flux quantum
is correct: $\Delta_{\Phi_0}(\nu =1/3) > \Delta_{\Phi_0}(\nu =2/5) >
\Delta_{\Phi_0}(\nu =3/7)$ 
\cite{Ambrumenil}.  Note that we obtained for particle 
excitations:
$\Delta_{p}(\nu =1/3) > \Delta_{p}(\nu =3/7) >
\Delta_{p}(\nu =2/5)$ (data not presented). Finally, due to particle
hole symmetry within the lowest  lLL
identical gaps at $\nu=2/3$, $\nu=3/5$, and $\nu=4/7$ were observed.
ii) At $\nu = 1/3$, and in the spirit of the Laughlin wave function,
removing one flux quantum is equivalent to adding a quasiparticle. 
Since a full electron is 
composed of three quaisiparticles, one expects if the 
interaction between quasiparticles is small:  $\Delta_{\Phi_0}(\nu=1/3)
\sim \Delta_{p}(\nu=1/3)/3$. This is confirmed in figure 2. However, it
appears that $3\Delta_{\Phi_0}(\nu=1/3)$ is slightly smaller than
$\Delta_{p}(\nu=1/3)$ which 
support the existence of an attractive interaction between the
quaisiparticles. Hence, the quasiparticle picture seems to be
justified also on the lattice.
iii) The gap at  $\nu=1/2$ shows very large fluctuations as
a function of system size. 
From our numerical data, we clearly may not obtain reliable information 
on the nature (incompressible or fermi liquid
\cite{Halperin})
of the $\nu=1/2$ ground state for the considered short range interactions.

Figure 3 plots the  density-density  
correlation functions (\ref{den-den}) for the $\nu=1/3$
incompressible state on a $24 \times 24$ lattice. The lattice
results (solid circles) are compared to the prediction of the Laughlin
wave function \cite{Note1} (solid line).  For this lattice size, $l_0/a \sim 2$ and it is
remarkable to see that there is already nearly perfect agreement between
the lattice and continuum results. Figure 3 also plots the 
quaisiparticle $g(r)$ which is obtained by removing a flux quantum from
the $\nu=1/3$ state. $g(r)$ for the quaisiparticle state is also a
smooth function at small distances, and as expected decreases 
for $r\rightarrow 0$ with
a smaller power law than the $\nu=1/3$ state.

The numerical data show explicitly that the Hamiltonian $H_0 + H_I$ is scale
invariant at the numerically accessible odd denominator filling
fractions. 
The lattice constant is an irrelevant length scale,
and the characteristic correlation functions scale with $l_0$, the only
length scale involved.
Hence, the choice of the interaction (\ref{H_I}) is a 
renormalization group fixed point.

Mean field theories for partially
filled lLL at $\nu = \frac{1}{2m+1}$ can be obtained by
decomposing the electron operators into an odd number of fermionic
partons with charges $\frac{e}{2m+1}$,  $\psi_{r}=\prod_{\alpha =1}^m
\psi^{\alpha}_{r}$, subjected to the
constraints ${\psi^{\alpha}_{r}}^{\dagger}\psi^{\alpha}_{r}=
{\psi^{\beta}_{r}}^{\dagger}\psi^{\beta}_{r} \>,\> \forall \alpha,\beta$
\cite{Wen}.
If each species fills up the lLL with $N_\Phi
= \frac{N}{2m+1}$ the mean field decoupling of the product of parton
operators is stable if the interaction between the electrons is
repulsive and short ranged \cite{Wen}. The parton wave functions have
$2m+1$ components when we put them onto a torus. On one hand, this
implies that the product wave function is periodic and hence is a
candidate for an incompressible electron state \cite{Jain}.
On the other hand the
constraints on the parton currents induce fractional statistics
\cite{Wen}. The important point is that in our approach the correct
multivaluedness of anyons on a torus \cite{Einarsson} is automatically
provided.
The hierarchy of IQL states, which we partially observe on the
lattice, can be understood analogously to the continuum situation. The
continual condensation of localized quasiparticle excitations into
Laughlin states relies at the mean field level on the self-similarity of the
Hofstadter spectrum. This property assures
that the levels at smaller magnetic
field strengths, which are populated by quasiparticle excitations, have
the same overall structure. The continuum one particle spectrum is
trivially self-similar, since in this case
the energies depend linearly on the
magnetic field strength. By the introduction of the lattice the
hierarchy of IQL states is {\em unfold}.

To conclude, we have considered tight-binding spinless 
electrons on a square lattice
in a strong magnetic field and submitted to a next-neighbor repulsive 
interaction.  We defined lattice Landau levels (lLL) as well as a continuum 
limit which insures that they scale to continuum Landau levels.
By projecting onto the lowest lLL and already on 
lattices where $l_0/a \sim 2$ the FQHE at odd filling fractions 
was observed with remarkable precision.
Those observations open a new region,
the continuum edge of the Hofstadter spectrum, where the FQHE
may be observed and studied.

We are very indebted to A. Muramatsu who suggested this approach to
the FQHE.
We would like to thank W.Hanke, M. Imada, C. K\"ubert,  K. Kusakabe,
F. V. Kusmartsev and S. Meixner
for motivating discussions. F.F.A. would like to thank
the DFG for 
financial support under the grant number  Ha 1537/6-2 
as well as the Japanese Ministry of Education 
and S.B. the  financial support of the 
BMFT under the grant number 03-HA3WUE.

\newpage
\subsubsection*{Figure captions}
\newcounter{bean}
\begin{list}%
{Fig. \arabic{bean}}{\usecounter{bean}
                   \setlength{\rightmargin}{\leftmargin}}

\item  The magnetic length  as a function of the lattice size.
An estimate of $l_0$ on the lattice (symbols) is obtained by fitting 
the lattice value of the density-density correlation functions
to their  continuum value
(\ref{den-den}) in the case of a filled first landau level \cite{Note2}.
We consider two lattice 
directions, $ \vec{r} = r(1,0)$ (downward triangles) and 
$ \vec{r} = r(1,1)$ (circles) as well as $m=1,3 \and 5$. 
The solid lines correspond to the continuum value: $l_0/L = \sqrt{m}/
\sqrt{2 \pi N} $ where $N$ corresponds to the linear size of the lattice.

\item Gaps to adding a full electron  $\Delta_p(\nu)$ (\ref{Gapp}) 
as well as  a flux quanta $\Delta_{\Phi_0}(\nu)$
(\ref{Gapph}) 
for various fillings and as a function of the lattice size.
The data refers to the special case $m=1$ (see equation (\ref{flux})).

\item  Density-density  correlation functions for the
$\nu = 1/3$ state on a $24\times 24$ lattice (solid circles). 
The open circles plot $g(r)$ for the quaisiparticle on a $23\times 23$ 
lattice. The solid line corresponds to $g(r)$ as obtained from the Lauglin
wave function at $\nu=1/3$ \cite{Note1}. 
The data refers to the special case $m=1$ (see equation (\ref{flux})).
	
\end{list}

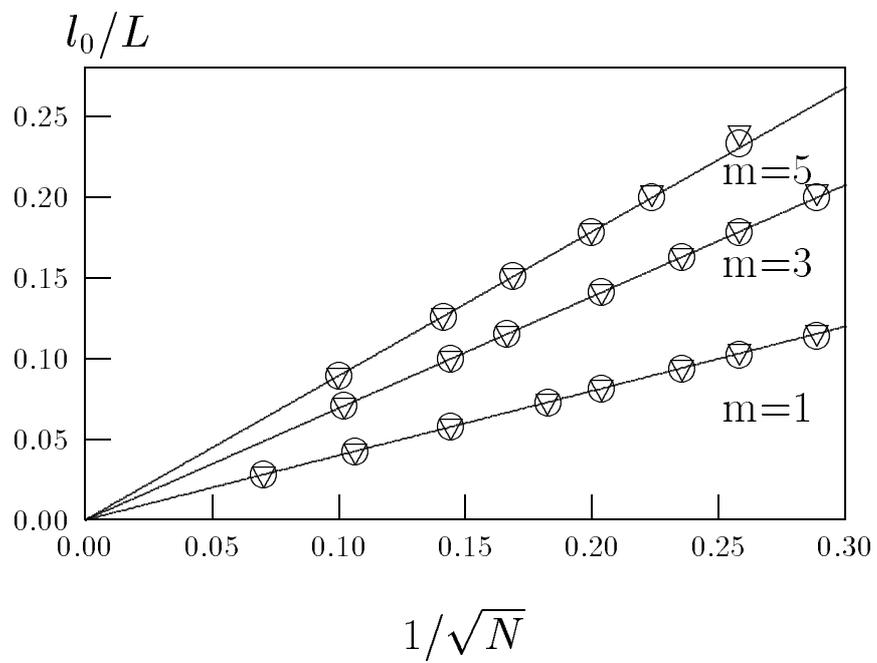
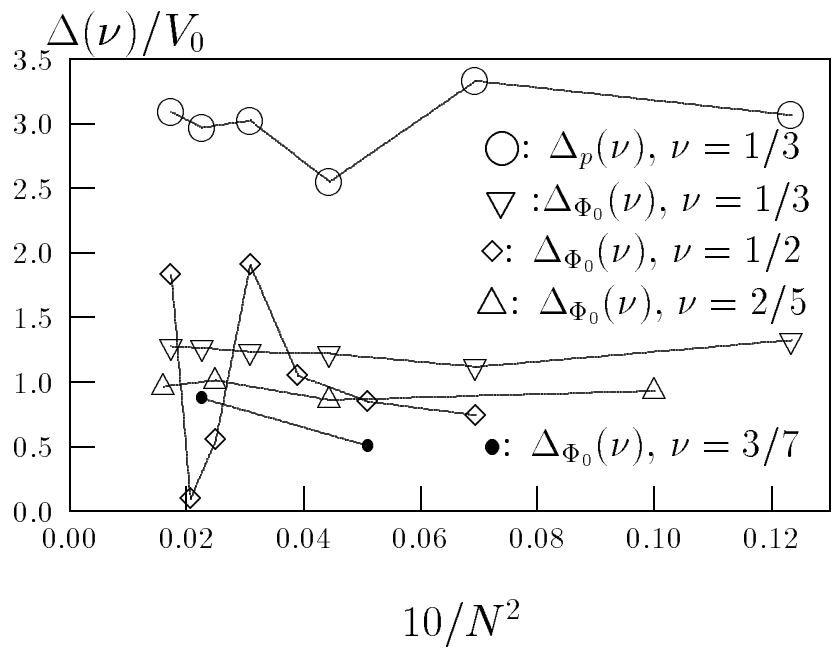
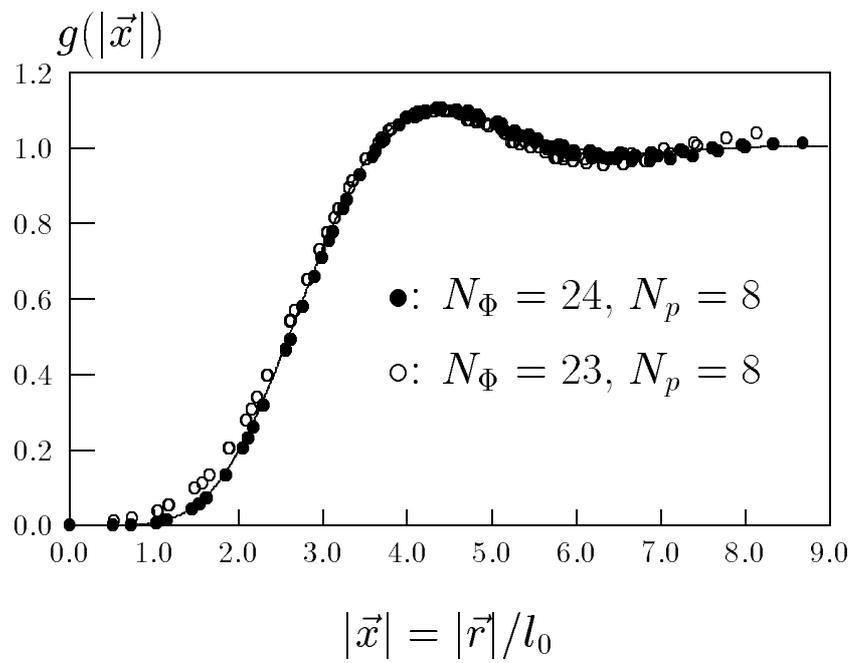
\end{document}